\documentclass[12pt]{article}
\usepackage{graphicx}
\begin{document}

\rightline{DFTT 66/98}
\rightline{LYCEN 98103}

\begin{center}

{\bf \Large The additive quark model revisited:\\ hadron and photon
 induced cross-sections}

\medskip
{\bf P. Desgrolard}({\footnote{E-mail: desgrolard@ipnl.in2p3.fr}}),
{\bf M. Giffon}({\footnote{E-mail: giffon@ipnl.in2p3.fr}}),
{\bf E. Martynov}({\footnote{E-mail: martynov@bitp.kiev.ua}}),
{\bf E. Predazzi}({\footnote{ E-mail: predazzi@to.infn.it}}).

\bigskip
($^{1,2}$){\it
Institut de Physique Nucl\'eaire de Lyon, IN2P3-CNRS et Universit\'{e}
Claude Bernard,\\
43 boulevard du 11 novembre 1918, F-69622 Villeurbanne Cedex, France\\}

($^3$){\it
N.N. Bogoliubov Institute for Theoretical Physics, National Academy of
Sciences of Ukraine,
252143, Kiev-143, Metrologicheskaja 14b, Ukraine\\}

($^4$){\it
Dipartimento di Fisica Teorica - Universit\`a di Torino
and Sezione INFN di Torino, Italy}

\end{center}

\bigskip
\bigskip
\noindent
{\bf \large Abstract}
The standard additive quark model and the ensuing counting rules are
modified to take into account the quark-gluonic content of the Pomeron and of
the
secondary Reggeons.  The result is that a much improved
description of $pp, \pi p, \gamma p$ and $\gamma \gamma$ cross-sections is
achieved.

\vskip 1.cm
\section{Introduction}

The Additive Quark Model (AQM)\cite{LevFra,LipShe,KokLVH,Kok} has provided for a
long time a simple and successful model to describe, in particular, the
main relations between the high energy cross-sections of different hadronic
processes \cite{AnShSh,Mart}.  Considering for example the pion-nucleon and
the nucleon-nucleon interactions, one finds that the relation
$\sigma_{tot}^{\pi N}/\sigma_{tot}^{NN}=2/3$ is in agreement with the available
experimental data within an accuracy of a few percent.  A linear dependence
of the amplitudes on the number of quarks inside the scattered hadrons was
confirmed on more fundamental grounds through QCD-like models
\cite{GunSop,LevRys}.

An interesting case to which we can apply (and test) the AQM lies into extending
it to photon induced
reactions because the data on these processes are now available up to quite high
energies ($\sqrt{s}\approx 200$ GeV for $\gamma p$ and
$\sqrt{s}\approx 100$ GeV for $\gamma \gamma$ inelastic cross-sections)
\cite{e1,e2}.  The three processes
($pp,\ \gamma p$ and $\gamma \gamma$) are related via
unitarity and factorization and this is the only set of related
processes for each of which we have data.
For the hadronic $pp-,\ \pi p-,\ \pi \pi-$reactions, the data on
$\pi \pi-$interaction are in fact absent and $\pi p$ total
cross-sections are known only up to relatively low energies
($\sqrt{s}<30$ GeV).

Considering the above mentioned processes, first we show (Sect.~2) that the
standard AQM does not describe the data with sufficiently high quality
\footnote{We cannot compare the quality of our fit with those
presented in some recent papers \cite{bl,do} because their $\chi^2$
is not given. Differences in their and our predictions for
cross-sections at higher energies are discussed in Sect.~3}.
Next, In Sect.~3, we propose a modified AQM that takes into account the
quark-gluonic content of the
exchanged Reggeons. As a first try, we take into account Pomeron and $f$-Reggeon
because they contribute to all amplitudes.  The suggested
modification provides a much improved agreement with the
experimental data.

In order to make clear the content of our modification we will not consider here
the scattering processes at $t\neq 0$.  The parameterization of the scattering
amplitudes at $t\neq 0$ is much more complex than at $t=0$.  It will be the
subject of a forthcoming paper.

\section{The old additive quark model}

\subsection{The Pomeron}

The traditional additive quark model treats the elastic scattering of two
hadrons at high energy as  Pomeron exchange between two quarks, one in each
hadron.  From the
point of view of the quark-gluon picture, the Pomeron is represented by a gluon
ladder with end points coupled with quark lines.  The simplest diagram
describing the main contribution
to elastic hadron-hadron amplitude in the old AQM is exemplified in Fig.~1.

\vskip 0.7cm
\begin{center}
\includegraphics*[scale=.7]{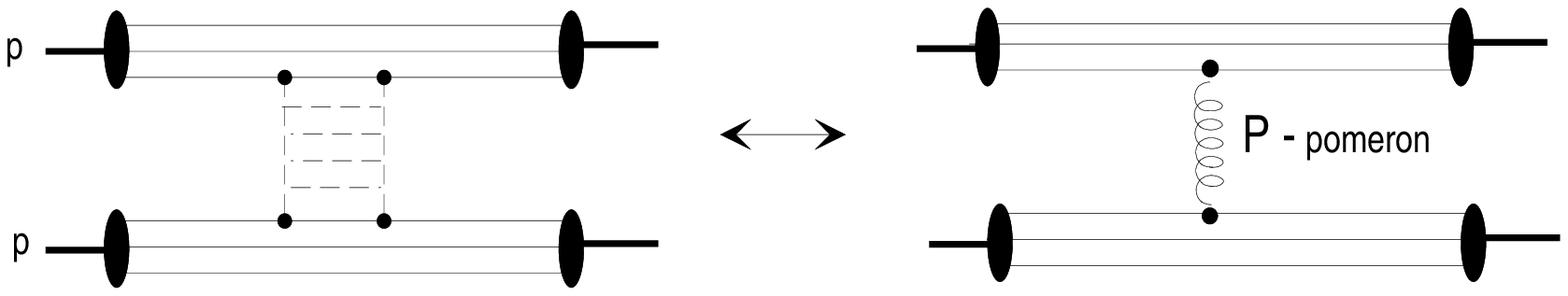}
\vskip 0.3cm
{\bf Fig.~1.} The Pomeron diagrams for $pp-$scattering in the traditional
additive quark model.
\end{center}

In accordance with AQM, when hadrons $h_1$ and $h_2$ (made of $n_1$ and
$n_2$ quarks) are colliding, the Pomeron contribution to the elastic amplitude
has the form
\begin{equation}\label{1}
A_{\cal P}^{(h_1h_2)}(s,t)=n_1n_2P_{h_1}P_{h_2}A^{(qq)}_{\cal
P}(s_{h_1h_2},t) G_{\cal P}^{(h_1)}(t)G_{\cal P}^{(h_2)}(t)
\end{equation}
where $\sqrt{P_{h_i}}$ is the probability of finding the hadron $h_i$ as a quark
system, $A^{(qq)}_{\cal P}(s,t)$ is the amplitude of elastic scattering of
quarks due to the Pomeron and the squared energy $s_{h_1h_2}$ will be defined
more precisely below ((9)). $G_{h_{i}}(t)$ is the form factor of the hadron
$h_i$; it takes into account a redistribution of momenta of the quarks inside a
hadron after the interaction of one of them with the Pomeron (each system of
quarks
should be preserved, after the interaction, as a hadron of the same kind).  It
is clear
that $G_{\cal P}^{(h_1)}(0) = G_{\cal P}^{(h_2)}(0)=1$ at $t=0$. In what
follows we apply the traditional and the modified AQM to describe the total
cross-sections
\begin{equation}\label{2}
\sigma_{tot}(s)=8\pi \Im mA(s,0)
\end{equation}
and the ratios of the real to the imaginary forward amplitudes
\begin{equation}\label{3}
\rho (s)={\Re eA(s,0)\over\Im mA(s,0)}.
\end{equation}

For the Pomeron contribution to the quark-quark scattering, we will
consider two sche\-mes. The first one is the
Supercritical Pomeron (SCP) ({\it i.e.} a Pomeron with an intercept larger than
one, a
variant of the Donnachie-Landshoff Pomeron (DLP)\cite{DL}, but
with a constant term added to reflect preasymptotic properties (this is nothing
but a simple pole in the complex angular momentum plane with unit intercept)
\begin{equation}\label{4}
A^{(qq)}_{\cal P}(s,0)=ig_1^2[-\zeta + (-is/s_0)^{\alpha_{\cal
P}(0)-1}]\ ,
\end{equation}
where $s_0=1$ GeV$^2$.

The second model is the Dipole Pomeron (DP) model (see, for instance
\cite{Bert}),
corresponding to a sum
of a simple pole and a double $j$-pole with unit intercepts
\begin{equation}\label{5}
A^{(qq)}_{\cal P}(s,0)=ig_1^2[-\zeta+\ln(-is/s_0)]\ .
\end{equation}
In both previous expressions the parameter $\zeta$ is expected to be
positive (from the fits to hadronic and $\gamma p$ cross-sections
\cite{dglm,dlm}).

As shown in \cite{dlm2}, the Pomeron contribution at $t\neq 0$ is more
complicate than (1) because each term in (4) and (5) should be
multiplied by {\it a priori} different vertex functions $G(t)$. Thus, at $t\neq
0$, (1) must
be rewritten as
\begin{equation}\label{6}
A_{\cal P}^{(h_1h_2)}(s,t)=n_1n_2P_{h_1}P_{h_2}\sum_{i=1,2}A^{(qq)}_{{\cal
P}i}(s_{h_1h_2},t) G_{{\cal P}i}^{(h_1)}(t)G_{{\cal P}i}^{(h_2)}(t),\
\end{equation}
where, generalizing (4) and (5),
\begin{equation}\label{7}
A_{{\cal P}1}^{qq}(s,t)=-ig_1^2\zeta (-is/s_0)^{\tilde \alpha_{\cal P}(t)-1},
\quad A_{{\cal P}2}^{qq}(s,t)=-ig_1^2L(s,t), \quad \tilde \alpha_{\cal P}(0)=1
\end{equation}
and
\begin{equation}\label{8}
L(s,t)=(-is/s_0)^{\alpha_{\cal P}(t)-1}, \qquad \alpha_{\cal P}(0)>1
\qquad \mbox{for SCP},
\end{equation}
$$
L(s,t)=\ln(-is/s_0)(-is/s_0)^{\alpha_{\cal P}(t)-1}, \qquad \alpha_{\cal P}(0)=1
\qquad \mbox{for DP}\ .
\eqno(8')$$
Generally speaking, the trajectories $\alpha_{\cal P}(t)$ and
$\tilde \alpha_{\cal P}(t)$  can differ not only by their intercepts but also by
their slopes.

At high energy, in the c.m. system, each hadron has the energy
$\sqrt{s/2}$ and, according to the AQM, each quark inside the hadron $h_i$ has
the energy $\sqrt{s/2}/ n_i$ (if $n_i$ is the number of quarks comprised in the
hadron $h_i$). Thus, the energy of each pair of quarks (one
from the hadron $h_1$ and the other from the hadron $h_2$) is
\begin{equation}\label{9}
  s_{h_1h_2}=\bigg (\frac{\sqrt{s/2}}{n_1}+\frac{\sqrt{s/2}}{n_2}\bigg )^2-
  \bigg (\frac{\sqrt{s/2}}{n_1}-\frac{\sqrt{s/2}}{n_2}\bigg )^2=
  {s\over n_1n_2}.
\end{equation}

There are nine (3$\times$3) similar diagrams in $pp$ scattering,
contributing to the corresponding amplitudes
\begin{equation}\label{10}
A^{(pp)}_{{\cal P}i}(s,t)=9 P^2_p\, A^{(qq)}_{{\cal P}i} (s/9,t)
(G^p_{{\cal P}i}(t))^2 , \qquad i=1,2\ .
\end{equation}
Considering also elastic $\pi p$-scattering with 3$\times $2 diagrams one
can write
\begin{equation}\label{11}
 A^{(\pi p)}_{{\cal P}i}(s,t)=6 P_pP_\pi \, A^{(qq)}_{{\cal P}i} (s/6,t)
 G^p_{{\cal P}i}(t)G^\pi_{{\cal P}i}(t), \quad i=1,2.
\end{equation}

Let us focus now on $\gamma$ induced processes for which the main Pomeron
contributions (in the AQM) are shown in Fig.~2.

\vskip 0.7cm
\begin{center}
\includegraphics*[scale=.7]{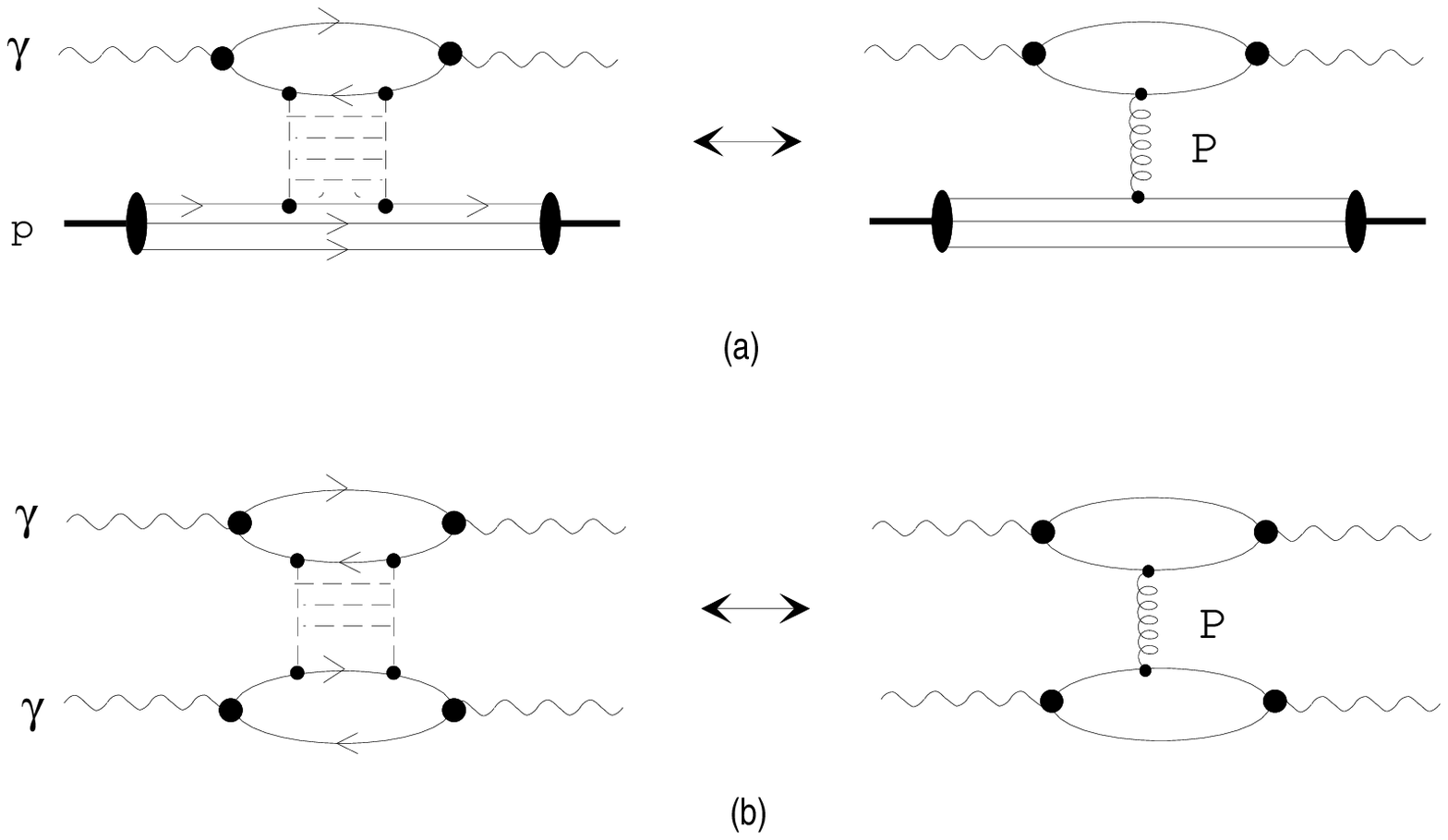}
\vskip 0.3cm
{\bf Fig.~2.} The Pomeron diagrams for $\gamma p$ (a) and $\gamma
\gamma $ (b) scattering in the old AQM.
\end{center}

\noindent
The simplest approximation
that describes the $\gamma p$ elastic scattering as due to the Pomeron is
\begin{equation}\label{12}
A^{(\gamma p)}_{{\cal P}i}(s,t)=6 \alpha P_p\, A^{(qq)}_{{\cal P}i} (s/6,t)
G^p_{{\cal P}i}(t)G^{\gamma }_{{\cal P}i}(t), \quad i=1,2\ ,
\end{equation}
where $\alpha=e^2/4\pi\approx 1/137$ is the fine structure constant.
(12) takes into account the $\gamma q\bar q$ vertices in the $q\bar q$ loop
at the upper block of the diagram of Fig.~2a.  Similarly, the relevant
$\gamma\gamma$
amplitude (Fig.~2b) has the form
\begin{equation}\label{13}
A^{(\gamma \gamma )}_{{\cal P}i} (s,t)=4 \alpha^2\,A^{(qq)}_{{\cal P}i} (s/4,t)
(G^{\gamma }_{{\cal P}i}(t))^2, \quad i=1,2\ .
\end{equation}

It is, however, more realistic to consider a different picture for
$\gamma p$
and $\gamma \gamma$ diagrams. In accordance with the Vector
Meson Dominance (VDM) model, the photon is transformed into a vector meson
which, after interacting with the Pomeron, comes back to a photon
state. Thus, we replace $\alpha\ \to\ P_\gamma$ in (12),(13)
where $\sqrt{P_{\gamma }}$ describes the transition of a $\gamma $ into a
pair $q\bar q$ (for instance, via a vector meson).

\subsection{Secondary Reggeons}
At the presently attainable (subasymptotic) squared energy $s$, beside the
Pomeron, one
should retain
also the contribution of other Reggeons ($f, \rho , \omega $ etc)
to the elastic amplitudes. It is usually assumed that they are
added to the Pomeron so that the
amplitude becomes
\begin{equation}\label{14}
A^{(h_1h_2)}(s,t)= A_{\cal
P}^{(h_1h_2)}(s,t)\ +\ n_1n_2P_{h_1}P_{h_2}\sum\limits_{R}
A^{(qq)}_{R}(s_{h_1h_2},t)G_{R}^{(h_1)}(t)G_{R}^{(h_2)}(t) ,
\end{equation}
where the Pomeron amplitude is detailed in the preceding section and the sum
over $R$ runs over all Reggeons contributing to the given process.  In
what follows, we will consider $pp,\ \pi p,\ \gamma p$ and $\gamma \gamma$
scattering at $\sqrt{s}\ge 4$ GeV and $t=0$.  Therefore, only $f$ and $\omega$
will contribute to $p^{\mp}p$ processes (here and in what follows $p^-\equiv
\bar p,\ p^+\equiv
p$), $f$ and $\rho $ to $\pi^{\mp}p$ and $f$ to $\gamma p$ and $\gamma
\gamma $. For the secondary Reggeons we take the standard form
\begin{equation}\label{15}
A^{(qq)}_f(s,t)= ig_f^2\bigg (-i{s/s_0}
\bigg )^{\alpha_f(t)-1}\ ,
\end{equation}
\begin{equation}\label{16}
A^{(qq)}_\omega (s,t)= g_\omega ^2\bigg (-i{s/s_0} \bigg )^{\alpha_\omega
(t)-1}\ ,
\end{equation}
\begin{equation}\label{17}
A^{(qq)}_\rho(s,t)=g_\rho^2\bigg (-i{s/s_0} \bigg )^{\alpha_\rho(t)-1}\ .
\end{equation}

\subsection{Complete AQM amplitude}

Collecting all the previous results, the relevant amplitudes at $t=0$ \footnote{
This is all we need for total cross-sections.} in
the old additive quark model for the four cases under investigation are
\begin{equation}\label{18}
  A^{p\bar p \choose pp}(s,0)=
9P^2_p\left[A_{\cal P}^{(qq)}(s/9,0)+A_{f}^{(qq)}(s/9,0)\pm
A_{\omega}^{(qq)}(s/9,0)\right],
\end{equation}
\begin{equation}\label{19}
  A^{\pi^-p \choose \pi^+p}(s,0)=
6P_\pi P_p\left[A_{\cal P}^{(qq)}(s/9,0)\pm
A_{\rho }^{(qq)}(s/6,0)\right],
\end{equation}
\begin{equation}\label{20}
A^{(\gamma p)}(s,0)=6P_\gamma P_p\left[A_{\cal P}^{(qq)}(s/6,0)
+A_{f}^{(qq)}(s/6,0)\right],
\end{equation}
\begin{equation}\label{21}
A^{(\gamma \gamma)}(s,0)=4P_\gamma^2\left[A^{(qq)}_{\cal
P}(s/4,0)+A_{f}^{(qq)}(s/4,0)\right]\
\end{equation}
where the Pomeron quark-quark amplitude $A^{(qq)}_{\cal P}$ is defined by (4) or
(5) and the Reggeon quark-quark amplitudes $A^{(qq)}_R$ are written in
(15)-(17).

\subsection{Comparison of the data with the old AQM}
The above amplitudes have been fitted to the experimental data
\cite{e1,e2,e3,e4} at $\sqrt{s}\geq 4$ GeV (totally 434 points) listed in
Table 1.

We did not include in our data set a few points on
$\rho^{\pi^\pm p}$ because of their large errors. They do not lead to any
noticeable change in the values of parameters and in the behaviour of the
curves.

Without any loss of generality we can take $P_p=1$ in the previous equations
since this acts as an overall multiplication parameter in the fit.

We compare three models of Pomeron:
\begin{itemize}
\item[a)]
DLP~: Supercritical Pomeron with $\zeta=0$ in (4) (which is close to the
Pomeron in \cite{DL}).
\item[b)]
SCP~: Supercritical Pomeron with free $\zeta $.
\item[c)]
DP~: Dipole Pomeron with $\alpha_{\cal P}(0)=1$.
\end{itemize}

\begin{table}\label{T1}
\caption{Number of experimental data points used in the fit to the
cross-sections and $\rho $-values of the various processes}
\medskip
\begin{center}
\begin{tabular}{|l|c|c|c|c|c|c|c|c|}
\hline
Observable &  $\sigma_{pp}$ & $\sigma_{\bar pp}$ & $\sigma_{\pi^- p}$
& $\sigma_{\pi^+ p}$ & $\sigma_{\gamma p}$ & $\sigma_{\gamma \gamma }$
& $\rho_{pp}$ & $\rho_{\bar pp}$ \\
\hline
Number of points & 85 & 51 & 49 & 83 & 68 & 17 & 64 & 17\\
\hline
\end{tabular}
\end{center}
\end{table}

The description of these data in all models are comparable to each other; the
$\chi^2$
for cases b) and c), $\chi^2/d.o.f.\approx 3.04$, is very close to that of case
a) $\chi^2/d.o.f.\approx 3.09$. It
is interesting to note, nevertheless, that, if the parameter
$\zeta $ is allowed to be free, the intercept of the Supercritical Pomeron
tends to 1 and the other parameters approach those obtained in the Dipole
Pomeron model. The same situation was observed in \cite{dglm}, where these
models were compared with all the data on the meson-nucleon and nucleon-nucleon
cross-sections and $\rho $-s. We will come back to these questions below when
discussing the modified AQM. In Figs.~3,4 we present the curves corresponding
to the Dipole Pomeron (case c)). The curves for both variants of SCP are very
close to the DP curves.

\vskip 0.7cm
\begin{center}
\includegraphics*[scale=.8]{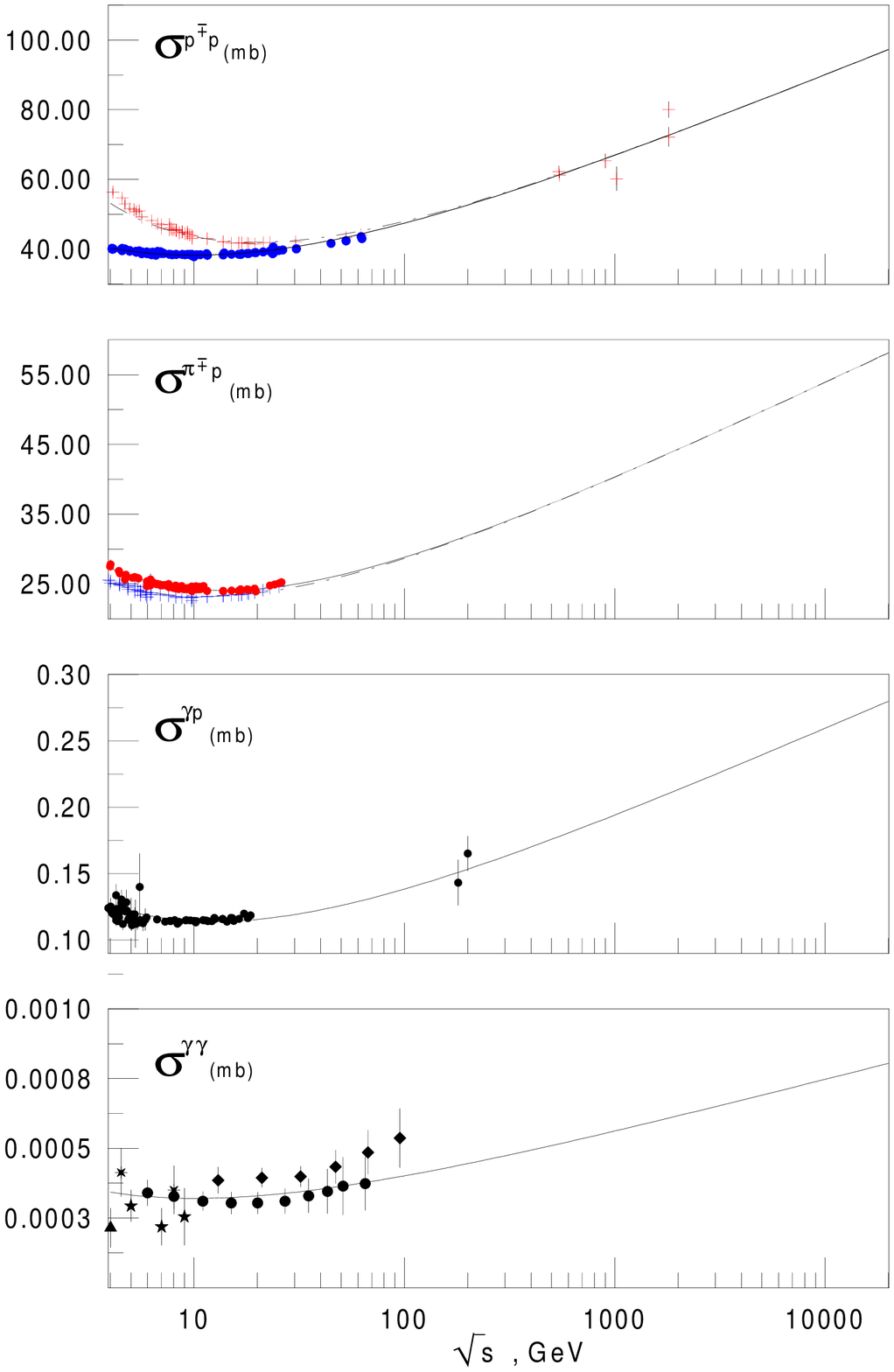}
\vskip 0.3cm
{\bf Fig.~3.} Total cross-sections described in old AQM with the Dipole Pomeron.
\end{center}

\vskip 0.7cm
\begin{center}
\includegraphics*[scale=.6]{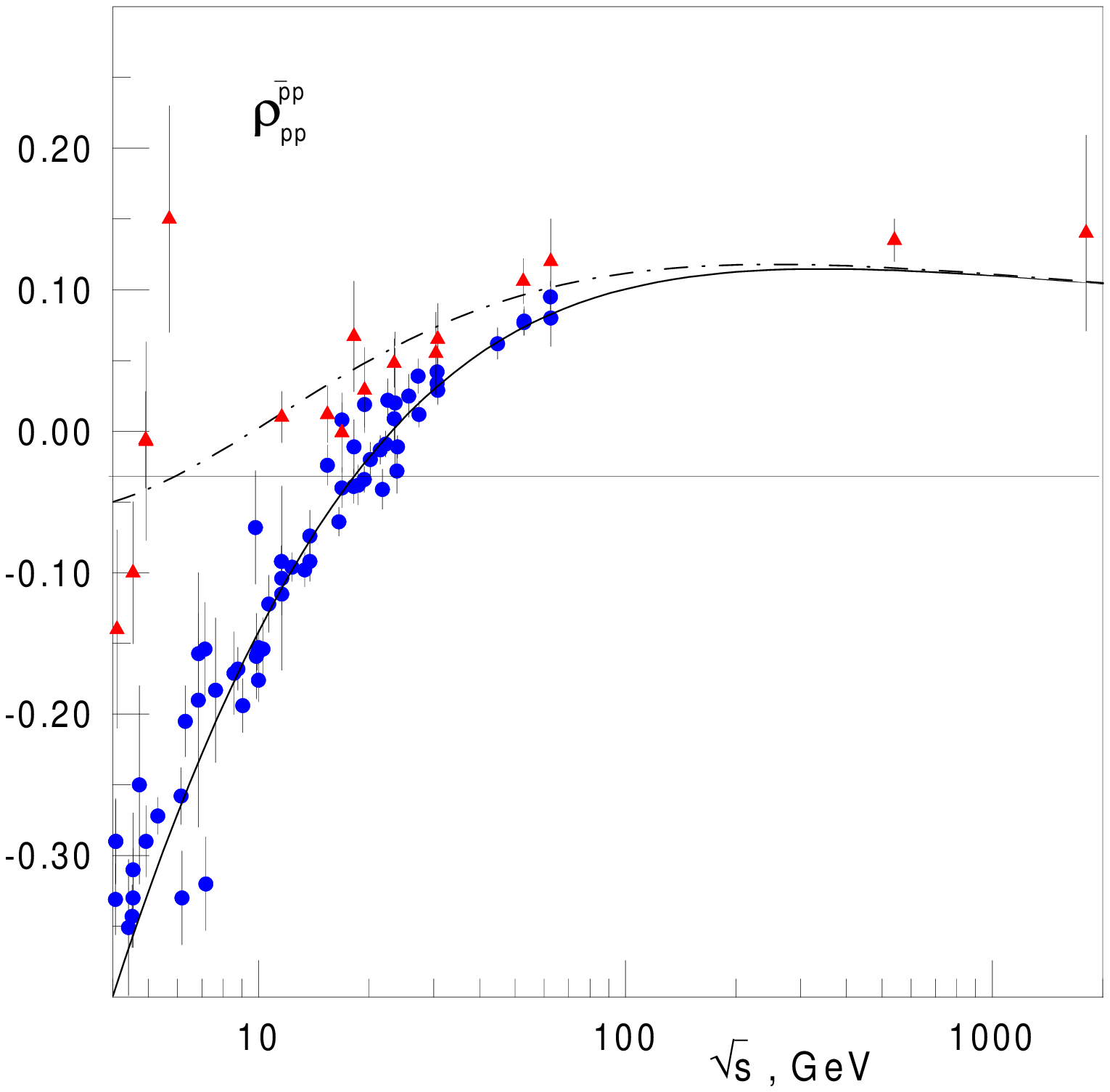}
\vskip 0.3cm
{\bf Fig.~4.} Ratio of the forward real to imaginary part for $pp$ and $\bar pp$
scattering  in the old AQM with the Dipole Pomeron.
\end{center}

\section{Modification of the additive quark model}

\subsection{Pomeron}
In the spirit of the QCD-like picture, the Pomeron (a
gluonic ladder as a first approximation) has at least four gluonic
edges coupled with quark lines. Besides the diagrams of Fig.1,2 however,
additional terms may contribute to the amplitudes. They correspond to diagrams
in which the Pomeron line is coupled with two quark lines rather
than with one only, leading to a modified additive quark model (MAQM). Examples
of such diagrams are shown in Fig.~5.
The cheapest price to pay for this generalization is an additional coupling
constant describing the vertex Pomeron - two quark lines. This
constant can be determined from the fit to experimental data and
can be considered as a measure of the deviation between the new counting
rules and the old ones.

 We would like to note here that in spite of an apparent analogy, the right
hand side Pomeron diagrams in Fig.~5 (as well as in the other Figures) are not
exactly the ladder diagrams, shown on the left hand side of Fig.~5.  The latter,
in fact, assume that each vertex of the gluonic ladder of the Pomeron couples to
the hadron via lines of individual quarks only
while the diagrams on the right hand side of Fig.~5 do not exclude that a soft
Pomeron can be exchanged between single quarks and pair of quarks (not
necessarly a
diquark; all possible states of this pair of quarks are "hidden" in the new
coupling constant) or between two pairs of quarks.
\vskip 0.7cm
\begin{center}
\includegraphics*[scale=0.8]{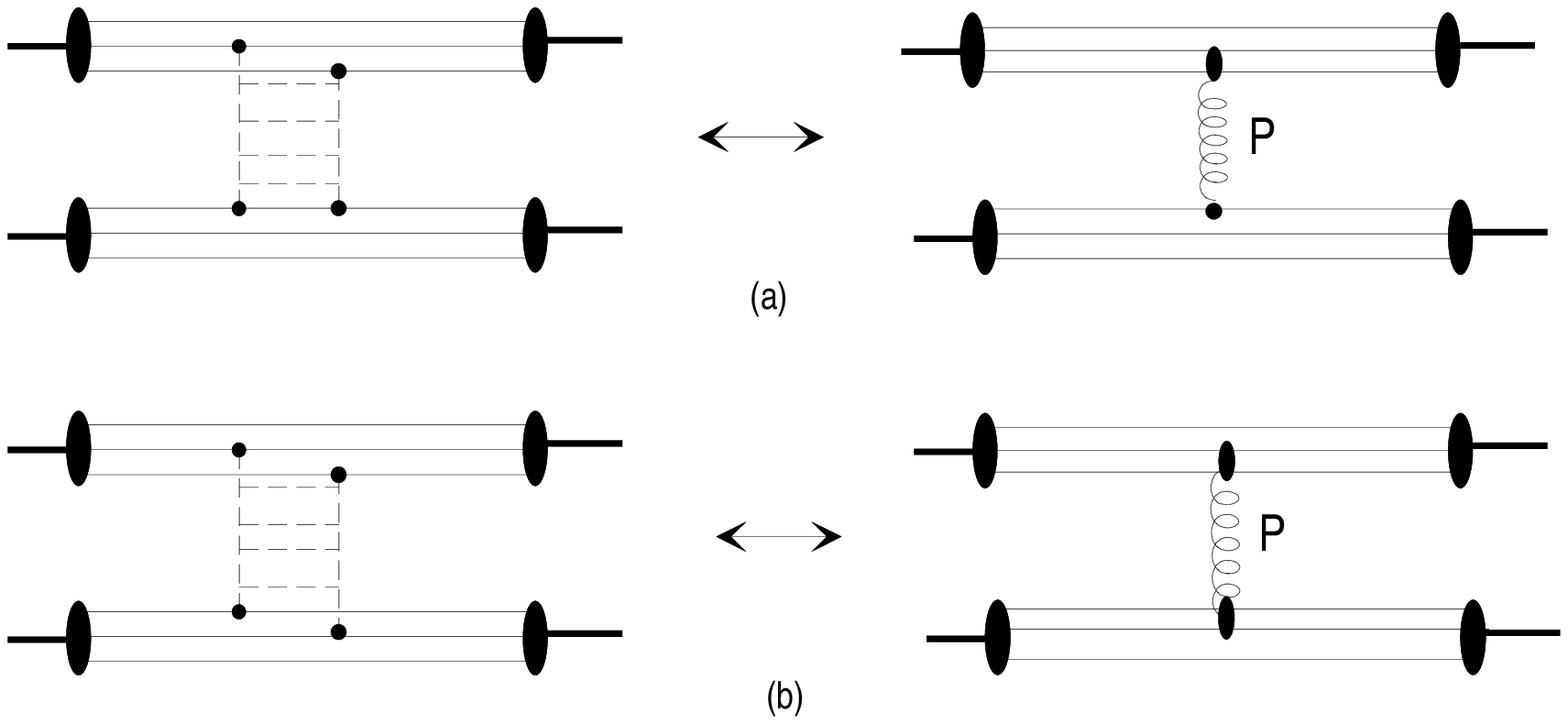}
\vskip 0.3cm
{\bf Fig.~5.} Additional Pomeron diagrams in the modified AQM (examples).
\end{center}

As it is well known, a soft Pomeron satisfying unitarity does not reduce to
ladder diagrams.  There is a difference between a two-gluon approximation to the
Pomeron (used for example in \cite{GunSop,LevRys}) and a soft Pomeron which is
certainly a more complicate object than a gluonic ladder.  Because a
calculable scheme for a soft Pomeron is not known, we are forced to rely
on phenomenological models.

Taking into account all possible diagrams that can contribute to each case, we
redefine the Pomeron contributions (10-13) in the following form (once again,
it is sufficient to write
all amplitudes at $t=0$ because our modification concerns the
counting rules rather than the form of the amplitudes; a generalization to
$t\neq 0$
is, however, immediate from the previous Section)

$p^\mp p$-interaction:
\begin{equation}\label{22}
  A^{(pp)}_{\cal P}(s,0)=9P_p^2[A^{(1)}_{\cal P}(s/9,0)+2A^{(2)}_{\cal P}
  (2s/9,0)+
  A^{(3)}_{\cal P}(4s/9,0)],
\end{equation}

$\pi^\mp p$-interaction:
\begin{equation}\label{23}
 A^{(\pi p)}_{\cal P}(s,0)=3P_\pi P_p[2A^{(1)}_{\cal P}(s/6,0)
  +3A^{(2)}_{\cal P}(s/3,0)+A^{(3)}_{\cal P}(2s/3,0)],
\end{equation}

$\gamma p$-interaction:
 \begin{equation}\label{24}
 A^{(\gamma p)}_{\cal P}(s,0)=3P_\gamma P_p[2A^{(1)}_{\cal P}(s/6,0)
  +3A^{(2)}_{\cal P}(s/3,0)+A^{(3)}_{\cal P}(2s/3,0)],
 \end{equation}

$\gamma \gamma $-interaction:
\begin{equation}\label{25}
  A^{(\gamma \gamma)}_{\cal P}(s,0)=P_\gamma^2[4A^{(1)}_{\cal P}(s/4,0)
  +4A^{(2)}_{\cal P}(s/2,0)+
  A^{(3)}_{\cal P}(s,0)]\ ,
\end{equation}
where we have defined
\begin{equation}\label{26}
  A^{(1)}_{\cal P}(s,0)=i[-\eta_1^2+g_1^2L(s)],
\end{equation}
\begin{equation}\label{27}
  A^{(2)}_{\cal P}(s,0)=i[-\eta_1\eta_2+g_1g_2L(s)],
\end{equation}
\begin{equation}\label{28}
  A^{(3)}_{\cal P}(s,0)=i[-\eta_2^2+g_2^2L(s)]\ .
\end{equation}
Here $L(s)=(-is/s_t)^{\alpha_{\cal P}(0)-1}$ in the SCP model and
$L(s)=\ln(-is/s_0)$ in the DP model (and $\eta_1,\, \eta_2$ are constants).

Each Pomeron term $A^{(i)}_{\cal P}(s,0),\ i=1,3$ is the sum of
two contributions~: the first corresponds to a double $j$-pole (with
couplings $g_1, g_2$ describing the vertices with one and
two quarks as indicated in Fig.~1 and Fig.~5) and the second
corresponds to a simple pole (with couplings $\eta_1$ and
$\eta_2)$. As already noted in Sect.2.1 a negative contribution of the
simple pole is suggested from fitting the hadronic
amplitudes to the data; this is why we have a negative sign in front of the
$\eta$'s in
(26)-(28). The available data, however, are not sufficient to determine four
coupling constants ($g_k, \eta_k$); for this reason we consider the
simpler case in which
$$ \eta_1^2/g_1^2=\eta_2^2/g_2^2=\zeta .  $$

As follows from unitarity, the total cross-sections for $nn, \pi n$
and $\pi
\pi$ interactions (by cross-section $nn$ and $\pi n$ we mean here
$\sigma_{nn}=(\sigma_{pp}+\sigma_{\bar pp})/2$ and $\sigma_{\pi
n}=(\sigma_{\pi^+p}+\sigma_{\pi^-p})/2$) should satisfy  at asymptotic energies
the factorization relation \cite{Col}
$$\sigma_{\pi n}^2=\sigma_{\pi \pi }\sigma_{nn}.  $$
One can check that this relation holds also in the MAQM if
the constant terms in (26)-(28) are neglected. But the well known
relation, $\sigma_{\pi n}/\sigma_{nn}=3/2$, does not hold exactly  in the MAQM.
For the
Dipole Pomeron ($L(s)=\ln(-is/s_0)$) this is modified into
$$
\sigma_{\pi n}/\sigma_{nn}=\sigma_{\pi \pi}/\sigma_{\pi n}\approx
\frac{2P_\pi }{3P_p}(1-\frac{1}{2}\frac{g_1}{g_2}),
$$
if $g_1/g_2\ll 1$ as it is expected (and confirmed by data, see below).
The same relations are valid (under the replacement $\pi \rightarrow \gamma $ in
the
indices) for $nn,\, \gamma n$ and $\gamma \gamma $ processes.

\subsection{Secondary Reggeons.}
While the Pomeron is mostly a gluonic state, which can be coupled
with any quark independently of its flavour, the Reggeons must be
considered as $q\bar q$-states (see Fig.~6). Concerning the $f$-Reggeon (as
well as other Reggeons with vacuum quantum numbers) which, being neutral is a
mixing of $u\bar u$ and $d\bar d$, not of $u\bar d$ and $d\bar u$ states (we
ignore here the small
contribution of other $q\bar q$ states to the $f$-meson and,
consequently, to the $f-$Reggeon), its diagrammatic structure within our
approach
is shown in Fig.~6.

Thus, for the $pp$-diagram the $f$-Reggeon can couple only to
quarks with identical flavour. There are four diagrams where the
$f$-Reggeon couples to $u$-quarks and one to $d$-quarks. The summation leads to
5
$f-$Reggeon diagrams in $pp$-scattering instead of 9 diagrams for
the Pomeron and for the $f$-Reggeon in the old AQM. One obtains

\begin{equation}\label{29}
  A^{(pp)}_f(s,0)=5P_p^2A^{(qq)}_f(s/9,0).
\end{equation}

\vskip 0.7cm
\begin{center}
\includegraphics*[scale=.7]{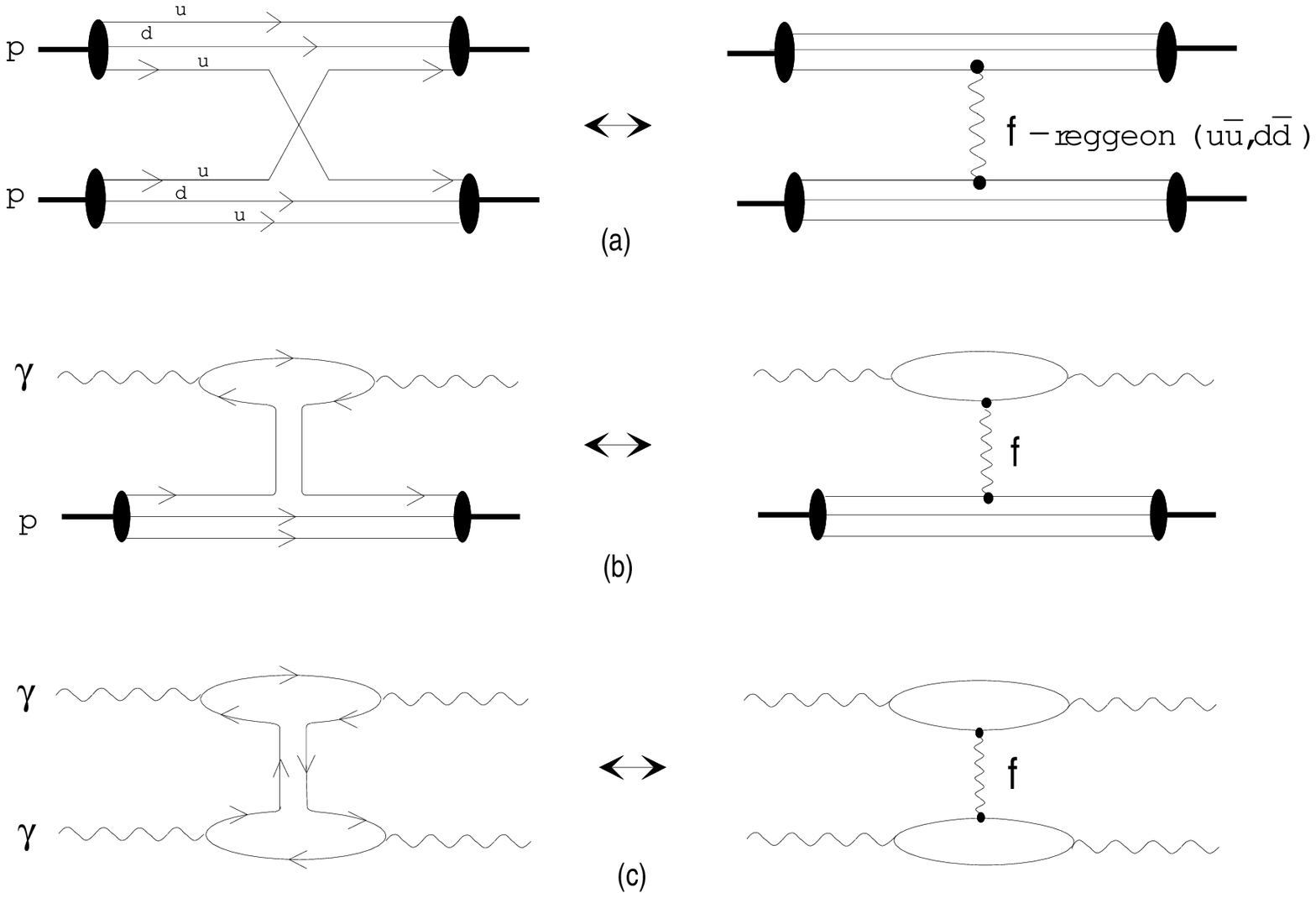}
\vskip 0.3cm
{\bf Fig.~6.} The $f-$Reggeon diagrams for $pp$ ,$\gamma p$ and
$\gamma
\gamma $ scattering.
\end{center}

Similarly, for  the $\pi p$ diagrams
\begin{equation}\label{30}
  A^{(\pi p)}_f(s,0)=3 P_{\pi} P_p A^{(qq)}_f(s/6,0).
\end{equation}
The same couplings apply to $\gamma p$ diagrams. The upper loop in Fig.~6b can
contain either
$u\bar u$ or $d\bar d$ quarks with 1/2 probability for each case.
Therefore, there are $2\cdot 2\cdot {1\over 2}=2$ terms for the
$u-$loop and $2\cdot{ 1\over 2}=1$ term for the $d-$loop leading to
\begin{equation}\label{31}
  A^{(\gamma p)}_f(s,0)=3P_\gamma P_pA^{(qq)}_f(s/6,0) .
\end{equation}
Performing a similar counting for the $\gamma \gamma $ amplitude, we
obtain
\begin{equation}\label{32}
  A^{(\gamma \gamma)}_f(s,0)=2P_\gamma^2A^{(qq)}_f(s/4,0) .
\end{equation}
The crossing-odd $\omega -$Reggeon contributes only to the $pp$ and $\bar pp$
amplitudes and we have
\begin{equation}\label{33}
A^{(pp)}_{\omega }(s,0)=5P_p^2A^{(qq)}_{\omega }(s/9,0).
\end{equation}
Similarly, the $\rho$ contribution to the $\pi^\mp p$ amplitudes
leads to
\begin{equation}\label{34}
A^{(pp)}_{\rho }(s,0)=3P_p^2A^{(qq)}_{\rho }(s/6,0).
\end{equation}

Strictly, we should consider two contributions since, in addition to the
previous
coupling with two quark lines with the same flavors ($uu$ or $dd$), we may also
have  another coupling with two quark lines with different flavors ($ud$),
because all
secondary Reggeons with vacuum quantum numbers are mixed states of $u\bar u$ and
$d\bar d$ (we neglect harder flavors).  We can describe such a transition
from a $u\bar u$ state to a $d\bar d$ one by a new constant (see Fig.~7).

\vskip 0.7cm
\begin{center}
\includegraphics*[scale=.8]{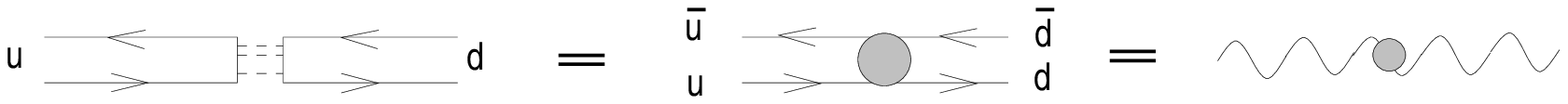}
\vskip 0.3cm
{\bf Fig.~7.} Mixing of $u\bar u$ and $d\bar d$ states in a
secondary Reggeon.
\end{center}
For  simplicity, we assume that this new contribution is given by multiplying
the old
$R-$Reg\-geon term by a constant $\lambda_R$. Note, however,  that the counting
rules for this new terms  are different
from those which couple identical quarks. Let us write down the complete
$f-$Reggeon contribution to the amplitudes.

$pp$-interaction:
\begin{equation}\label{35}
  A^{(pp)}_{f}(s,0)=P_p^2[5A^{(qq)}_{f1}(s/9,0)+4A^{(qq)}_{f2}(s/9,0)]=
  P_p^2(5+4\lambda_f)A^{(qq)}_f(s/9,0),
\end{equation}

$\pi p$-interaction:
\begin{equation}\label{36}
  A^{(\pi p)}_{f}(s,0)=3P_\pi P_p[A^{(qq)}_{f1}(s/6,0)
  +A^{(qq)}_{f2}(s/6,0)]=3P_\pi P_p(1+\lambda_f)A^{(qq)}_{f}(s/6,0),
\end{equation}

$\gamma p$-interaction:
\begin{equation}\label{37}
  A^{(\gamma p)}_{f}(s,0)=3P_\gamma P_p[A^{(qq)}_{f1}(s/6,0)
  +A^{(qq)}_{f2}(s/6,0)]=3P_\gamma P_p(1+\lambda_f)A^{(qq)}_{f}(s/6,0),
\end{equation}

$\gamma \gamma $-interaction:
 \begin{equation}\label{38} A^{(\gamma
\gamma)}_{f}(s,0)=2P_\gamma^2[A^{(qq)}_{f1}(s/4,0)
+A^{(qq)}_{f2}(s/4,0)]=2P_\gamma^2(1+\lambda_f)A^{(qq)}_{f1}(s/4,0)
\end{equation}
where
$$A^{(qq)}_{f1}(s,0)\equiv A^{(qq)}_{f}(s,0)$$
and
$$A^{(qq)}_{f2}(s,0)\equiv \lambda_f A^{(qq)}_{f}(s,0)\ .$$
The value of
$\lambda_f$ should be determined from a fit to the experimental
data.  Note that if $\lambda_f=1$ one comes back to the old counting rules for
the $f$ -Reggeon.

The $\omega$ and $\rho $ Reggeon contributions are easily derived from the above
expressions.  It should be noted that the counting rules for these Reggeons are
unimportant in our fit because we consider one by one the processes to which
they contribute.  Namely, $\omega $ contributes only to $pp$ and $\rho$
contributes only to $\pi p$ amplitudes.  Thus, it is sufficient to write them in
the old AQM form.

\subsection{Complete MAQM amplitudes}

Summarizing the results of the new counting rules, the final
expressions for the $t=0$ amplitudes of the reactions under investigation in the
Modified Additive Quark Model are:
\begin{enumerate}
  \item [1.]
  $pp$ and $\bar pp$ (or $p^\mp p$) amplitudes
\begin{equation}\label{39}
\begin{array}{rcl}
A^{p\mp p}(s,0) & = & P_p^2\{9[A^{(1)}_{\cal P}(s/9,0)+2A^{(2)}_{\cal
P}(2s/9,0) + A^{(3)}_{\cal P}(4s/9,0)]\\ & + & (5+4\lambda_f)A_f(s/9,0)\pm
9A_\omega (s/9,0)\},
\end{array}
\end{equation}
   \item[2.]
   $\pi^-p$ and $\pi^+p$  amplitudes
\begin{equation}\label{40}
\begin{array}{rcl}
 A^{\pi^\mp p}(s,0) & = & P_\pi P_p \{3[2A^{(1)}_{\cal
P}(s/6,0)+3A^{(2)}_{\cal P}(s/3,0)+A^{(3)}_{\cal
P}(2s/3,0)]\\ & + & 3(1+\lambda_f)A_f(s/6,0)\pm 6A_\rho (s/6,0)\},
\end{array}
\end{equation}
   \item[3.]
   $\gamma p$ amplitude
\begin{equation}\label{41}
\begin{array}{rcl}
A^{\gamma p}(s,0)&=&P_\gamma P_p\,\{3[2A^{(1)}_{\cal
P}(s/6,0)+3A^{(2)}_{\cal P}(s/3,0)
+A^{(3)}_{\cal P}(2s/3,0)]\\ &+&3(1+\lambda_f)A_f(s/6,0)\},
\end{array}
\end{equation}
  \item [4]
  $\gamma \gamma $ amplitude
\begin{equation}\label{42}
\begin{array}{rcl}
A^{\gamma \gamma }(s,0)&=&P_\gamma^2\,\{4A^{(1)}_{\cal
P}(s/4,0)+4A^{(2)}_{\cal P}(s/2,0)+A^{(3)}_{\cal P}(s,0)\\
&+&2(1+\lambda_f) A_f(s/4,0)\}.
\end{array}
\end{equation}
\end{enumerate}

\subsection{Comparison of the data with the Modified AQM}

We now proceed to utilize the same set of data used previously (Table 1) to
perform the same fit for the MAQM. The values of the free parameters for
the three models of
Pomeron considered are given in Table 2. It is evident that the Modified
Additive
Quark Model leads to a better description of the data:  the value of
$\chi^2/d.o.f.$
decreases from 3.04 to 1.78 for cases b) and c) and to 2.03 for case a).
 The behaviour
of $\sigma_{tot}$ and $\rho $ is shown in the Figs. 8,9 (once again, we confine
ourselves to plot the curves only for the case of the Dipole Pomeron).  The
improvement for $\sigma^{p^\mp p}$ and $\rho^{p^\mp p}$ is quite visible.
The improvement for $\pi p$ and $\gamma p$ is less clearly visible in the
figures but exists.

The first conclusion is, therefore, that the modified
AQM agrees with data noticeably better than the old AQM does.

We also note that the Supercritical Pomeron with an
additional constant term (i.e.  with $\zeta =\eta_1^2/g_1=\eta_2^2/g_2 \neq 0$
in (26-8) is very close to
the Dipole Pomeron.  As a matter of fact, given the small value of $\epsilon $
obtained from the fit (see Table 2), $\epsilon
=\alpha_{\cal
P}(0)-1\approx 0.0005$, one can write the Supercritical Pomeron contribution,
for instance to the $pp$
amplitude, in a form undistinguishable in practice from the Dipole Pomeron case
$$
A_{\cal P}^{pp}=ig_1^2[-\zeta +(-is/s_0)^\epsilon ]\approx ig_1^2[-\zeta
+1+\epsilon \ln(-is/s_0)]=i\tilde g_1^2[-\tilde \zeta +\ln(-is/s_0)]
$$
where $\tilde g_1^2=\epsilon g_1^2, \quad \tilde \zeta=(\zeta -1)/\epsilon$.
From the parameters given in Table 2, we find $\tilde g_1\approx
$0.31, $\tilde \zeta \approx $3.02 which are close to the corresponding
parameters of the Dipole Pomeron. The parameters of the other Reggeons are also
close to those obtained for the Dipole Pomeron model.

\vskip 0.7cm
\begin{center}
\includegraphics*[scale=.8]{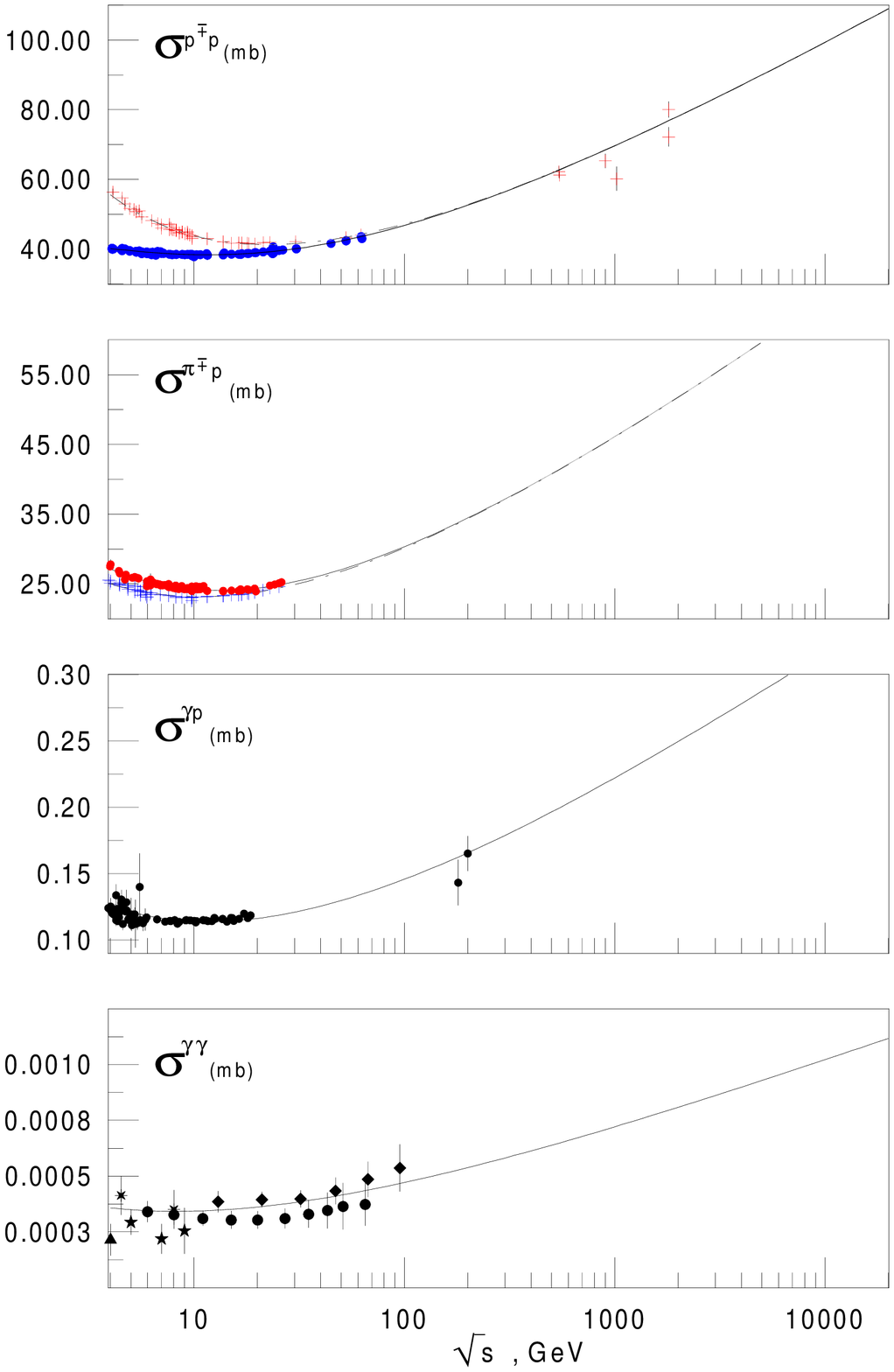}
\vskip 0.3cm
{\bf Fig.~8.} Total cross-sections described in the MAQM with the Dipole
Pomeron.
\end{center}

\vskip 0.7cm
\begin{center}
\includegraphics*[scale=.6]{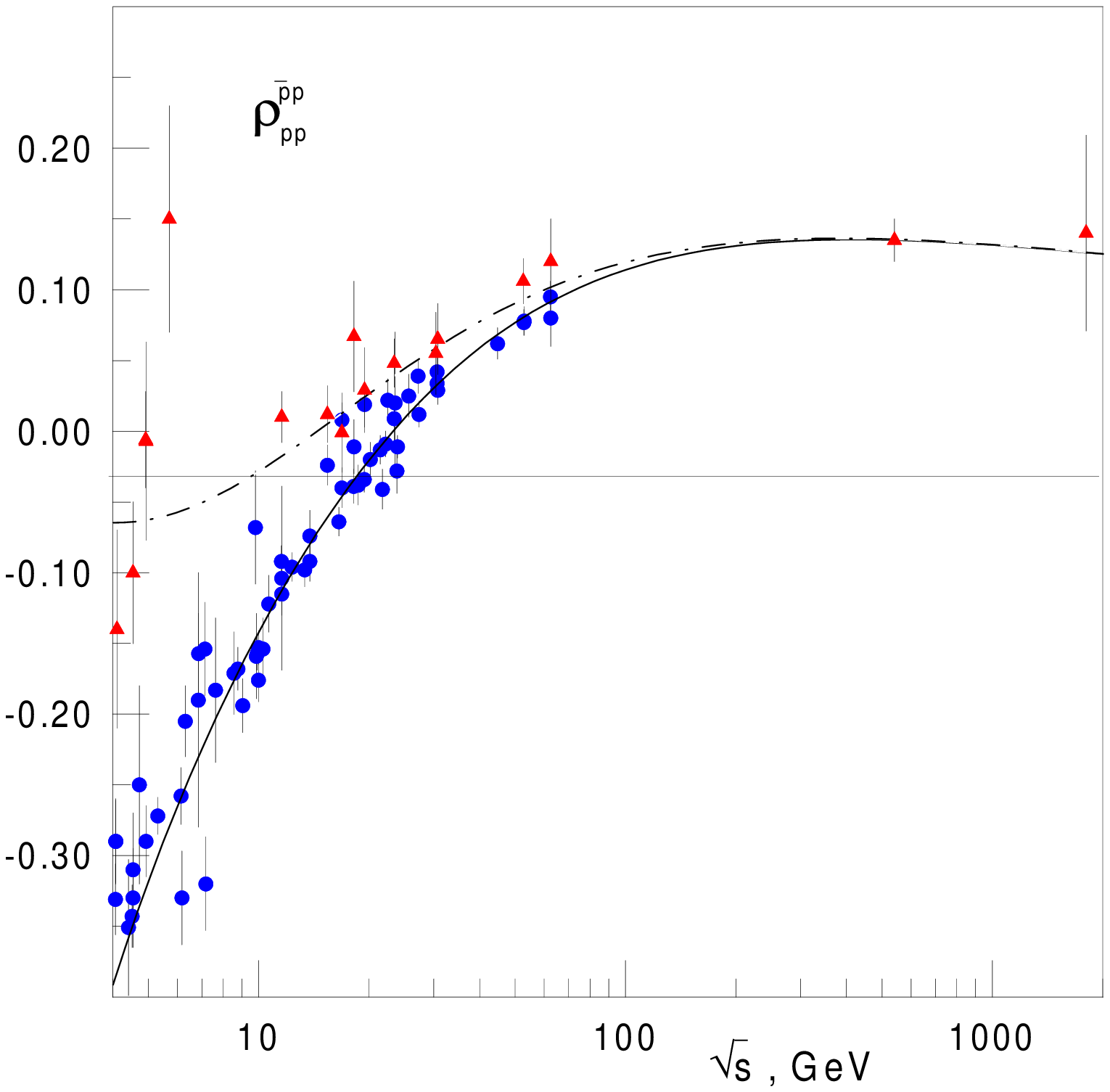}
\vskip 0.3cm
{\bf Fig.~9.} Ratio of the forward real to imaginary part for $pp$ and $\bar pp$
scattering  in the MAQM with the Dipole Pomeron.
\end{center}

\begin{table}\label{T2}
\caption{The values of parameters obtained in MAQM for three variants of Pomeron
.}
\medskip
\begin{center}
\begin{tabular}{|l|c|c|c|}
\hline
Parameters &  SCP, $\zeta =0$ & SCP, $\zeta \neq 0$ & DP\\
\hline
$g_1\ ({\rm  GeV}^{-1})$ & 0.583 & 13.346 & 0.317\\
$g_2\ ({\rm GeV}^{-1})$ & -0.079 & -0.826 & -0.024 \\
$\alpha_{\cal P}(0) $ & 1.101 & 1.0005 & 1.0\ (fixed) \\
$\zeta $ & 0.0\ {\rm (fixed)} & 1.003 & 3.399\\
$P_p $ & 1.0\  {\rm (fixed)}&1.0\ {\rm (fixed)} & 1.0\ {\rm (fixed)} \\
$P_\pi $ & 0.848  & 0.925 & 0.919 \\
$P_\gamma $ & 0.0041 & 0.0044 & 0.0044\\
$g_f\ ({\rm GeV}^{-1})$ & 0.822 & 1.120 & 1.112 \\
$\alpha_f (0) $ & 0.661 & 0.803 & 0.810 \\
$\lambda_f $ & 0.094 & 0.343 & 0.439 \\
$g_\omega\ ({\rm GeV}^{-1})$ & 0.396 & 0.395 & 0.395 \\
$\alpha_\omega (0) $ & 0.403 & 0.418 & 0.421 \\
$g_\rho\ ({\rm GeV}^{-1})$ & 0.230 & 0.221 & 0.222 \\
$\alpha_\rho (0) $ & 0.592 & 0.586 & 0.587 \\
\hline
$\chi^2/{\rm d.o.f}.$ & 2.025 & 1.784 & 1.780 \\
\hline
\end{tabular}
\end{center}
\end{table}

From this point of view, we can say that the Dipole Pomeron is preferable
to the Supercritical Pomeron.  From the theoretical point of view it will never
violate the Froissart-Martin unitarity bound, and from the phenomenological
point of view it has one parameter less (because $\alpha_{\cal
P}(0)=1$).

Our analysis of data does not support the conclusion drawn in \cite{bl} about
$\sigma_{inel}^{\gamma \gamma }$, that the preliminary OPAL are doubtful
and that the VMD selects the L3 data\footnote{The predictions of
Ref.\cite{bl} are based on the counting rules of the old AQM, which should be
modified as we have argued.}.  One can see from Fig.~8 that
the theoretical curve goes precisely between the points of these two groups.
Generally, we predict higher values of $\sigma^{\gamma p}$ and
$\sigma^{\gamma \gamma}$ than given in \cite{bl}, but smaller than those
obtained for these cross-sections in the mini-jets model \cite{cgp}.

\section{Conclusion}
Our main result is the following.  A proper account of the quark-gluonic content of Pomeron and $f$-Reggeon leads to slightly
modified
counting rules for the quark-quark amplitudes when constructing the
hadron-hadron, photon-hadron and photon-photon amplitudes.  The additional new
Pomeron
terms give $\approx $10\% of the whole Pomeron contributon, while for
$f$-Reggeon
the new term contributes $\approx $ 30\% of the whole $f$-Reggeon component.

The important role of these terms is confirmed by the analysis of the data on
the
total cross-sections of hadron and photon induced processes.  They lead to a
decrease of the $\chi^2/d.o.f.$ by approximately 40\%, thus improving the
description of the data.

In conclusion, we have shown that a modification of the additive quark model in
which one takes into proper account the contribution of more diagrams leads to a
quantitatively better fit of all available $t=0$ data. While not dramatic, this
improvement gives us hope that the MAQM will give a substantially better
result when the model will be applied outside $t=0$. This we plan to do in the
near future.
\vskip 0.5 cm
\noindent
{\bf Acknowledgements.} Financial support is gratefully acknowledged from the
IN2P3 of France and from the INFN and the MURST of Italy. E.M. wishes also to
thank
the Theory groups of the Universities of Lyon and Torino for their hospitality.

\end{document}